\documentclass[conference]{IEEEtran}
\IEEEoverridecommandlockouts
\newcommand{\etal}{\textit{et al.}}
\usepackage{cite}
\usepackage{amsmath,amsfonts}
\usepackage{algorithmic}
\usepackage{graphicx}
\usepackage{textcomp}
\usepackage{makecell}
\usepackage{colortbl}

\usepackage[hidelinks]{hyperref}
\usepackage{multirow}
\usepackage{graphicx,stackengine,scalerel}
\usepackage{amsmath}

\usepackage[dvipsnames]{xcolor}
\usepackage{amssymb}

\usepackage{array}

\newcolumntype{P}[1]{>{\centering\arraybackslash}p{#1}}

\newcommand{\greenArrow}{\textcolor{ForestGreen}{$\blacktriangle$}~}
\newcommand{\redArrow}{\textcolor{red}{$\blacktriangledown$}~}

\newcommand{\hlineThicc}{\noalign{\hrule height 1pt}}

\usepackage{array}
\newcolumntype{C}[1]{>{\centering\arraybackslash}p{#1}}

\def\BibTeX{{\rm B\kern-.05em{\sc i\kern-.025em b}\kern-.08em
    T\kern-.1667em\lower.7ex\hbox{E}\kern-.125emX}}
    \makeatletter

\newcommand{\linebreakand}{%
  \end{@IEEEauthorhalign}
  \hfill\mbox{}\par
  \mbox{}\hfill\begin{@IEEEauthorhalign}
}

\begin{document}

\title{Optimizing Deep Learning Models to Address Class Imbalance in Code Comment Classification}

\author{
    \IEEEauthorblockN{Moritz Mock}
    \IEEEauthorblockA{\textit{Faculty of Engineering} \\
    \textit{Free University of Bozen-Bolzano}\\
    Bolzano, Italy \\
    moritz.mock@student.unibz.it}
    \and
    \IEEEauthorblockN{Thomas Borsani}
    \IEEEauthorblockA{\textit{Faculty of Engineering} \\
    \textit{Free University of Bozen-Bolzano}\\
    Bolzano, Italy \\
    thomas.borsani@student.unibz.it}
    \linebreakand
    \IEEEauthorblockN{Giuseppe Di Fatta}
    \IEEEauthorblockA{\textit{Faculty of Engineering} \\
    \textit{Free University of Bozen-Bolzano}\\
    Bolzano, Italy \\
    giuseppe.difatta@unibz.it}
    \and
    \IEEEauthorblockN{Barbara Russo}
    \IEEEauthorblockA{\textit{Faculty of Engineering} \\
    \textit{Free University of Bozen-Bolzano}\\
    Bolzano, Italy \\
    barbara.russo@unibz.it}
}

\maketitle

\begin{abstract}
Developers rely on code comments to document their work, track issues, and understand the source code. As such, comments provide valuable insights into developers' understanding of their code and describe their various intentions in writing the surrounding code. Recent research leverages natural language processing and deep learning to classify comments based on developers' intentions. However, such labelled data are often imbalanced, causing learning models to perform poorly.
This work investigates the use of different weighting strategies of the loss function to mitigate the scarcity of certain classes in the dataset. In particular, various RoBERTa-based transformer models are fine-tuned by means of a hyperparameter search to identify their optimal parameter configurations. Additionally, we fine-tuned the transformers with different weighting strategies for the loss function to address class imbalances.
Our approach outperforms the STACC baseline by 8.9 per cent on the NLBSE’25 Tool Competition dataset in terms of the average F1$_c$ score, and exceeding the baseline approach in 17 out of 19 cases with a gain ranging from -5.0 to 38.2.
 The source code is publicly available at
\href{https://github.com/moritzmock/NLBSE2025}{https://github.com/moritzmock/NLBSE2025}.
\end{abstract}


\section{Introduction}
\label{sec:introduction}
Various artefacts, such as comments, commit messages, and issue tracker logs, describe developers' activity and intentions while writing their source code~\cite{ZampettiEtAl2021ComparisonMeans,MockEtAl2024developers}.
Among these, code comments are predominantly used, as developers often prefer them \cite{ZampettiEtAl2021ComparisonMeans}. Comments have different intentions, and recent research aims to classify \cite{PascarellaBacchelli2017} them in order to understand 
various characteristics of the developers' code (e.g., technical debt, vulnerability~\cite{RussoEtAl2022WeakSATD,MockEtAl2024MADE-WIC}). The distribution of comments across the various classification classes is typically imbalanced, with some classes being more represented than others (for example, \cite{bavota2016large}).  
Thus, the classification may turn poor and, as such, of little use. The NLBSE'25 challenge \cite{nlbse2025} provides a labelled dataset of comments for three programming languages and a baseline classification of them, STACC. 
\begin{table}[h!t]
    \centering
    \caption{Characteristics of the dataset; positive and negative instances for each comment Label, as well as the degree of imbalance towards the positive class.}
    \label{tab:dataset}
    \begin{tabular}{p{0.0005\textwidth}lccc}
                \hlineThicc
         & Label & positive & negative & positive\%\\ \hlineThicc
         \multirow{7}{*}{\rotatebox[origin=c]{90}{Java}} & Summary &4,502&4,837& 48.2\%\\
         & Ownership &312&9,027& 3.3\%\\
         & Expand &611&8,728& 6.5\%\\
         & Usage & 2,524 & 6,815 & 27.0\%\\
         & Pointer &1,088&8,251& 11.7\%\\
         & Deprecation &132&9,207& 1.4\%\\
         & Rational &379&8,960& 4.1\%\\ \hline
         \multirow{5}{*}{\rotatebox[origin=c]{90}{Python}}& Usage &699&1,591& 30.5\%\\ 
         & Parameters &700&1,590& 30.6\%\\
         & Development Notes &251&2,039& 11.0\%\\
         & Expand &407&1,883& 17.8\%\\
         & Summary &429&1,861& 18.7\%\\ \hline
         \multirow{7}{*}{\rotatebox[origin=c]{90}{Pharo}}& Key Implementations &221&1,366& 13.9\%\\ 
         & Example &666&921& 42.0\%\\
         & Responsibility &297&1,290& 18.7\%\\ 
         & Class Reference &50&1,537& 3.2\%\\
         & Intent &181&1,406& 11.4\%\\
         & Key Message &257&2,133& 16.2\%\\
         & Collaborators &86&1,501& 5.4\%\\ \hlineThicc
    \end{tabular}
\end{table}
Table \ref{tab:dataset} illustrates the distribution of the comments over labels and programming languages in the provided dataset. Also, in this case, we can see that the positive instances are underrepresented in all classes, but two (\texttt{Java - Summary} and \texttt{Pharo - Example}), with a percentage ranging from 1.4\% to 30.6\%. 

In this work, we aim to improve the classification performance of the baseline approach by using RoBERTa-based transformers \cite{LiuEtAl2019RoBERTa}, optimising their hyperparameters with fine-tuning, and incorporating loss functions weights to address the imbalance of positive instances.
We have considered four transformers: two pre-trained on an English corpus and two pre-trained on different code-related large-scale datasets. 
The comparison also allows us to discuss whether code-based pre-trained models outperform general-purpose pre-trained models in classifying code comments. To control for imbalanced data, we have further applied different weighting strategies for the loss function.
Overall, our approach outperforms the baseline approach, which is an adoption of the winner from two years ago \cite{AlKaswanEtAl2023STACC} leveraging a SetFit implementation \cite{TunstallEtAl2022SetFit}, with an average F1\textsubscript{c} score of 72.6, surpassing the 63.7 average F1\textsubscript{c} score from the baseline approach. For individual F1\textsubscript{c} score, the performance is improved in 17 classes out of 19 and up to 38.2 points in the F1\textsubscript{c}.

\subsection{Dataset}
\label{sec:dataset}

We leveraged the dataset provided by the challenge, which is a subset of the original dataset introduced by Rani \etal~\cite{RaniEtAl2021Dataset}. The dataset comprises comments, file names, and the label vector for three programming languages: Java, Python, and Pharo. Java and Pharo have seven distinct comment classes, while Python has five. Each comment is labelled for at least one class, and a comment can be labelled for multiple classes. For each of the programming languages, the number of available instances varies: 9,339, 2,290, and 1,587, respectively, for Java, Python, and Pharo.  
Furthermore, the dataset provided for each language is already split into \textit{train and test} subsets (80\%-20\%). In this research, we use the data attributes \texttt{combo} and \texttt{labels} of the dataset without modifications. 

\section{Approach}
\label{sec:approach}
\begin{figure}[ht]
    \centering
\includegraphics[width=\columnwidth]{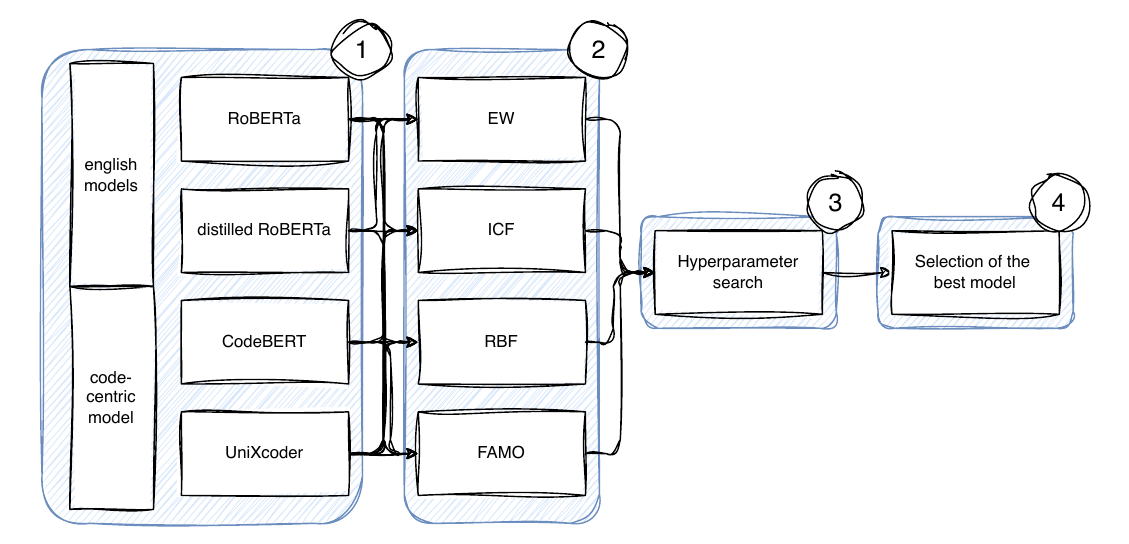}
    \caption{Overview of the approach}
    \label{fig:approach}
\end{figure}

Figure \ref{fig:approach} illustrates the steps of our approach: (1) model exploration, (2) definition of the loss functions, (3) hyperparameter search, and (4) best model selection.
\subsection{Model Exploration}
The first step of our methodology explores ``data hungry'' transformer models based on the BERT-Architecture, such as RoBERTa~\cite{LiuEtAl2019RoBERTa}, CodeBERT~\cite{FengEtAl2020CodeBERT}, UniXcoder~\cite{GuoEtAl2022UniXCoder}, and a distilled version of RoBERTa~\cite{SanhEtAl2019DistilBERT}.
RoBERTa, CodeBERT, and UniXcoder each have around 125M parameters, while the distilled version of RoBERTa has 82M parameters, making them highly expressive and powerful.
While RoBERTa is trained on an English corpus, CodeBERT and UniXcoder are trained on a corpus containing code; hence, we hypothesize that CodeBERT and UniXcoder may have superior performance due to the source code-related knowledge encoded during their pre-training. We fine-tune them for the downstream task of comment classification; fine-tuning refers to the task of adapting the weights of a pre-trained model for a given task. 

\subsection{Loss-function}
\label{sec:loss}
The loss function is used to calculate the penalty that the model will receive based on its performance during the training phase~\cite{WangEtAl2020Ranking}. The default loss function employed by our models is a linear combination of Binary Cross-Entropy losses (BCE), with one BCE for each class. Due to the class imbalance of this problem, equally weighted BCEs may suffer from a negative bias for underrepresented classes.
To tackle class imbalance, we have then considered alternative methods to differentiate the weights of BCE functions statically or adaptively during the fine-tuning of the model as described in the following.

\paragraph{Inverse Class Frequency (ICF)}
The loss function is built by adding weights $w_c = \frac{1}{f_c}$ to the BCE for each class where $f_c$ is the frequency of the class $c$. 
This balancing strategy has the advantage of giving more value to the underrepresented classes. 

\paragraph{Ranking-Based Frequency (RBF)}
We have leveraged the balancing strategy of Wang \etal~\cite{WangEtAl2020Ranking}, for which the weights of the individual BCEs are the inverse ranked frequency of the classes, i.e., the frequency of the most predominant class in the dataset is the weight of the most underrepresented class. 

\paragraph{Adaptive weights based on FAMO}
Lastly, we have investigated the use of a dynamic weighting method, FAMO~\cite{LiuEtAl2024Famo}, during the fine-tuning process. This approach leverages a weighted computation of the different losses involved in the optimisation process. 
The weights are adjusted at each training batch based on the speed of convergence of each class, with losses converging slower receiving greater importance and losses converging faster being penalised. This strategy attempts to harmonise the convergence speed of all loss functions. We refer to this strategy as \textit{FAMO}.

\section{Experimental Setup}
\label{sec:setup}   

\subsection{Hyperparameter Search}
\label{sec:hyper-parametersearch}
We performed a hyperparameter search to identify the best combination of hyperparameters for each programming language, as shown in Table \ref{tab:hyper-parametersearch}. The highest average F1 determines the best combination of hyperparameters. The search space yielded 180 possible hyperparameter combinations. Each combination was tested across each programming language, resulting in three experiments run for each setup. 
For the FAMO strategy, three learning rates (25e-2, 25e-3, 25e-4) and three weight decay (1e-2, 1e-3, 1e-4) have been validated, resulting in an overall number of experiments of 1620, which corresponds to 180 x 3 (learning rates) x 3 (weight decay).
The values for the hyperparameter search have been selected based on literature \cite{FuEtAl2022linevul}. On the other hand, the upper limit for the batch size was defined on the average and median number of input tokens, which were 20 and 25, respectively, across the complete dataset such that we could ensure that truncation of the input did not happen during the learning or the prediction.

\begin{table}[t]
    \centering
    \caption{Search space for hyperparameters.}
    \label{tab:hyper-parametersearch}
    \setlength{\tabcolsep}{5pt}
    \begin{tabular}{l|c}
    \hlineThicc
        Parameter     & Search Space        \\ \hlineThicc
        Batch Size    & [1, 2, 4, 8, 16]    \\
        Epoch         & [1, 3, 5, 10]       \\
        Learning Rate & [3e-5, 4e-5, 5e-5]  \\
        Weight Decay  & [0, 0.01, 0.001]    \\ \hline
        loss weights*& EW, ICF, RBF, FAMO  \\ \hlineThicc
        \multicolumn{2}{l}{* Not a hyperparameter of the model, but influential to it.}
        
    \end{tabular}
\end{table}

\subsection{Metrics and Implementation Details}
\label{sec:metrics}
The competition provided a set of metrics for the evaluation of the approach performance for each of the categories.
The metrics are precision ($P_c$), recall ($R_c$), and F1 ($F1_c$), which are calculated for each category. Lastly, for the overall scoring, the average F1$_c$ of all categories is considered.
\begin{equation*}
    P_c = \frac{TP_c}{TP_c+FP_c},~R_c = \frac{TP_c}{TP_c+FN_c},~F1_c = 2*\frac{P_c*R_c}{P_c+R_c}
\end{equation*}
The experiments are conducted on six nodes consisting of a Nvidia A100 GPU with 80GB VRAM and 192 GB of RAM in a server with the processor Xeon 4208 with 16 cores. 
The code was implemented leveraging pytorch 2.5.1 and transformers 4.35.0. Furthermore, we provide the top-performing models through Hugging Face and a Colab Notebook for testing purposes; both can be found in the replication package~\cite{MockEtAl2025Replication}.

\section{Results}
\label{sec:results}
We have run all our models with different combinations of hyperparameters and weighting strategies for the loss function on the three datasets.   
Columns 2-4 of Table \ref{tab:results} illustrate performance indices for the best transformer according to the average F1\textsubscript{c} over each language class and loss weights strategy. The last column on the right, finally, reports the performance values for the best model with the best loss weights strategy. The $\Delta$ F1\textsubscript{c} shows the increment or decrement with respect to the baseline approach (STACC).  
\paragraph{EW}
Leveraging a RoBERTa-based transformer-based model already outperforms the baseline average F1\textsubscript{c} by 5.2\% in the worst-case scenario. While selecting the best model and combination of hyperparameters for each of the programming languages, the performance is further increased by 2.4\% in the average F1\textsubscript{c}. EW refers to equal weights for the loss function.
\paragraph{ICF}
Applying the inverse class frequency to the loss function (BCE) further increases the overall performance of the best-performing models by 0.2\%.
\paragraph{RBF}
The ranking-based frequency further increased the performance to 72.1\% (0.6\% from the previous step), outperforming the baseline in 17 out of 19 classes. And even further improving the performance of the category \texttt{Java - Deprecation} of the ICF approach by 8.6\% in the F1\textsubscript{c}.
\paragraph{FAMO}
The FAMO approach did not perform as expected. We observed that F1\textsubscript{c} performed well for the more populated classes, while the model was underperforming for others. This might be caused by the high scarcity of the dataset, which prevented the optimisation strategy from learning properly. In fact, for the classes with lower sample sizes, the F1\textsubscript{c} score was found to be close to zero. Consequently, we did not surpass an F1\textsubscript{c} of 60, and therefore, the results of this approach were excluded in Table \ref{tab:results}.
 
\paragraph{Best Solution}
We have selected the models with the best F1\textsubscript{c} for the best hyperparameters and any loss function. The hyperparameter values can be obtained from Table \ref{tab:hyper-parametersearch}. For \texttt{Pharo} and \texttt{Python}, the ICF approach resulted in the best programming language specific average F1\textsubscript{c}, while for \texttt{Java} was the RBF approach. Lastly, the best performing fine-tuned pre-trained model in our settings was always CodeBERT \cite{FengEtAl2020CodeBERT}. The hyperparameters which resulted in the best combination were for the batch size 2 for Python and 4 for Java and Pharo, the number of epochs was always 10, the learning rate was 4e-5 for Pyhton and 3e-5 for Java and Pharo, and the weight decay was 0 for Python, 0.01 for Pharo, and 0.001 for Java.

Overall, the results show an improvement in the performance over the baseline approach STACC and its model MiniML. 
CodeBERT appears to be the best model across the languages, indicating the benefit of pre-training on code-based datasets. Weights are important to control for imbalance, but the type of weights optimisation strategy may depend on the language. 
The results also show one consistent and negative behaviour. For the classification of comments labelled \texttt{Java - Deprecated}, the baseline approach outperforms all our best models. Looking at Table~\ref{tab:dataset}, we see that the class is the most unrepresented with the fewest instances. We believe that the transformers we used require a good amount of data to be effective. Thus, other techniques that manipulate the sample to balance the class (e.g., adding synthetic data) may be more suitable for this case. 
The submission score, based on the formula provided by the NLBSE'25 Challenge \cite{nlbse2025}, is \textbf{0.44}, calculated from the average F1 = 0.726, average measured runtime of 11.6 seconds, and average measured GFLOPS of 155,300. 

\begin{table*}[h!t]
    \centering
    \setlength{\tabcolsep}{5.9pt}
    \caption{Comparison of our results with the baseline approach (STACC). $\Delta$ F1\textsubscript{c} indicates the difference between the best configuration and the baseline. The arrow shows whether the F1\textsubscript{c} improved or deteriorated with respect to the baseline. EW=Equal Weights, ICF=Inverse Class Frequency, RBF=Ranking-Based Frequency.}
    \label{tab:results}
    \begin{tabular}{p{0.0005\columnwidth}l|P{1.5em}P{1.5em}P{2.25em}||P{1.5em}P{1.5em}P{2.25em}|P{1.5em}P{1.5em}P{2.25em}|P{1.5em}P{1.5em}P{2.25em}|p{0.01px}@{\hspace{-4pt}}|P{1.5em}P{1em}P{2.25em}P{3em}}
    \hlineThicc 
    && \multicolumn{3}{c||}{\multirow{2}{*}{STACC}} &\multicolumn{9}{c|}{Loss weighting strategy}& &\multicolumn{4}{c}{\multirow{2}{*}{Best solution}}
    \\ \cline{6-14} 
    && && & \multicolumn{3}{c|}{\makecell{EW}}& \multicolumn{3}{c|}{\makecell{ICF}}& \multicolumn{3}{c|}{\makecell{RBF}}&& \multicolumn{4}{c}{}\\ \hlineThicc
    &Class label & P\textsubscript{c} & R\textsubscript{c} & F1\textsubscript{c} & P\textsubscript{c} & R\textsubscript{c} & F1\textsubscript{c} & P\textsubscript{c} & R\textsubscript{c} & F1\textsubscript{c} & P\textsubscript{c} & R\textsubscript{c} & F1\textsubscript{c} && P\textsubscript{c} & R\textsubscript{c} & F1\textsubscript{c} & $\Delta$ F1\textsubscript{c} \\ \hline
    && \multicolumn{3}{c||}{} & \multicolumn{3}{c|}{CodeBERT}& \multicolumn{3}{c|}{CodeBERT}& \multicolumn{3}{c|}{CodeBERT}&& \multicolumn{4}{c}{CodeBERT \& RBF}\\\hline
    \multirow{7}{*}{\rotatebox[origin=c]{90}{Java}}
    &Summary &87.3&82.9& 85.0&  90&90.9& \greenArrow90.5&88.6&92.2& \greenArrow90.3&90.2&91.1&\greenArrow90.7&&              90.2&91.1&\greenArrow90.7&~5.7\\ 
    &Ownership &100&100& 100&  100&100& ~~~100&100&100& ~~~100&100&100&~~~100&&                                              100&100&~~~100&~0.0\\ 
    &Expand &32.3&44.4& 37.4&  50.6&44.1& \greenArrow47.1&42.3&29.4& \redArrow34.7&50.5&50.0&\greenArrow50.2&&               50.5&50.0&\greenArrow50.2&12.8\\ 
    &Usage &91.1&91.8& 86.2&   93.6&87.6& \greenArrow90.7&90.4&87.2& \greenArrow88.8&92.6&87.0&\greenArrow89.7&&              92.6&87.0&\greenArrow89.7&~3.5\\ 
    &Pointer &73.8&60.0& 69.2& 81.9&98.4& \greenArrow89.4&82.5&97.3& \greenArrow89.3&81.0&97.3&\greenArrow88.4&&            81.0&97.3&\greenArrow88.4&18.9\\ 
    &Deprecation &87.3&82.9& 85.0&  91.7&73.3& \redArrow81.5&76.9&66.7& \redArrow71.4&100&66.7&\redArrow80.0&&               100&66.7&\redArrow80.0&-5.0\\ 
    &Rational &16.2&29.5& 20.9&  26.9&28.4& \greenArrow27.6&40.4&33.8& \greenArrow36.8&32.4&33.8&\greenArrow31.1&&           32.4&33.8&\greenArrow31.1&12.2\\ \hline
    &Java Average &69.7&70.2&69.1&76.4&74.7&\greenArrow75.2&74.4&72.4&\greenArrow73.1&78.1&75.1&\greenArrow\textbf{76.0}&&78.1&75.1&\greenArrow76.0&6.9\\
    \hline
    && \multicolumn{3}{c||}{} & \multicolumn{3}{c|}{CodeBERT}& \multicolumn{3}{c|}{CodeBERT}& \multicolumn{3}{c|}{CodeBERT}&& \multicolumn{4}{c}{CodeBERT \& ICF}\\\hline
    \multirow{5}{*}{\rotatebox[origin=c]{90}{Python}
    }&Usage &70.0&73.5& 71.7&  82.0&75.2& \greenArrow78.4&76.8&79.3&\greenArrow78.0&75.6&78.5& \greenArrow77.0&&             76.8&79.3&\greenArrow78.0&~6.3\\  
    &Parameters &79.3&81.2& 80.3&  88.5&84.4& \greenArrow86.4&87.4&81.2&\greenArrow84.2&84.0&85.9& \greenArrow84.9&&         87.4&81.2&\greenArrow84.2&~3.9\\  
    &Development Notes &24.3&48.7& 32.5&  38.6&41.5& \greenArrow40.0&42.2&46.3&\greenArrow44.2&47.2&41.5& \greenArrow44.2&&  42.2&46.3&\greenArrow44.2&11.7\\ 
    &Expand &43.3&76.5& 55.3&  62.1&57.7& \greenArrow59.8&59.3&54.7&\greenArrow56.9&54.5&52.0& \greenArrow53.2&&             59.3&54.7&\greenArrow56.9&~1.6\\ 
    &Summary &64.8&58.5& 61.5&  70.5&75.6& \greenArrow72.9&72.7&78.0&\greenArrow75.3&79.7&76.8& \greenArrow78.3&&            72.7&78.0&\greenArrow75.3&13.8\\  \hline
     &Python Average  &56.3&67.7&60.3&68.3&66.9&\greenArrow67.5&67.7&67.9&\greenArrow\textbf{67.7}&68.2&66.9&\greenArrow67.4&&67.7&67.9&\greenArrow67.7&7.5\\
    \hline
    && \multicolumn{3}{c||}{} & \multicolumn{3}{c|}{UniXcoder}& \multicolumn{3}{c|}{CodeBERT}& \multicolumn{3}{c|}{RoBERTa}&& \multicolumn{4}{c}{CodeBERT \& ICF}\\\hline
    \multirow{7}{*}{\rotatebox[origin=c]{90}{Pharo}}
    &Key Implementation &63.6&65.1& 64.3&  71.1&62.8& \greenArrow66.7&67.5&62.8& \greenArrow65.1&66.7&60.5& \greenArrow63.4&& 67.5&62.8&\greenArrow65.1&~0.8\\ 
    &Example &87.2&90.3& 88.7&  93.1&90.8& \greenArrow91.9&96.2&85.7& \greenArrow90.6&95.5&89.1& \greenArrow92.2&&            96.2&85.7&\greenArrow90.7&~2.0\\ 
    &Responsibility &59.6&59.6& 59.6&  55.4&59.6& \redArrow57.4&61.5&76.9& \greenArrow68.3&60.0&80.0& \greenArrow68.9&&       61.5&76.9&\greenArrow68.4&~8.8\\ 
    &Class References &20.0&50.0& 28.5&  75.0&75.0& \greenArrow75.0&60.0&75.0& \greenArrow66.7&66.7&60.0& \greenArrow63.2&&   60.0&75.0&\greenArrow66.7&38.2\\ 
    &Intent &71.8&76.6& 74.1&  73.5&83.3& \greenArrow78.1&90.0&90.0& \greenArrow90.0&82.0&79.3& \greenArrow80.6&&             90.0&90.0&\greenArrow90.0&15.9\\ 
    &Key Message &68.0&79.0& 73.1&  80.5&76.7& \greenArrow78.6&71.7&76.7&\greenArrow74.1&71.4&81.4& \greenArrow76.1&&         71.7&76.7&\greenArrow74.2&~1.1\\ 
    &Collaborators &26.0&60.0& 36.3&  33.3&60.0& \greenArrow42.8&50.0&60.0&\greenArrow54.5&74.2&45.8& \greenArrow56.6&&        50.0&60.0&\greenArrow54.5&18.2\\  \hline  &Pharo Average&56.6&68.7&60.7&68.8&72.6&\greenArrow70.1 & 71.0&75.3&\greenArrow\textbf{72.8}&73.8&70.9&\greenArrow71.5&&71.0&75.3&\greenArrow72.8&12.1\\
    \hlineThicc
    
    &\textbf{Grand Average} &61.4&69.0&63.7 &71.5&71.9& \greenArrow71.3&71.4&72.3& \greenArrow71.5&73.5&70.9& \greenArrow72.1&&            72.7&73.3&\cellcolor{gray!25}\greenArrow72.6 &~8.9\\ \hlineThicc
    \end{tabular}
\end{table*}

\section{Conclusions}
\label{sec:conclusion}
We employed four different pre-trained models, four different strategies in optimising the weights to the loss functions, and performed a hyperparameter search spanning 180 combinations.
We proposed a novel approach for code comment classification, which demonstrates significant advancement with respect to the baseline. By selecting CodeBERT as the model, implementing a hyperparameter search, and using different strategies in weighting the loss function, we have improved the classification performance across the three programming languages of the given dataset in all classification classes, but one. In this class, a data augmentation strategy should be considered (for instance, balancing classes with synthetic data). 
The superior performance of CodeBERT on comments' classification supports previous results at NLBSE'24~\cite{HaiEtAlDopamin}. 

\section*{Acknowledgments}
Moritz Mock is partially funded by the National Recovery and Resilience Plan (Piano Nazionale di Ripresa e Resilienza, PNRR - DM 117/2023). 
The work has been funded by the project CyberSecurity Laboratory no. EFRE1039 under the 2023 EFRE/FESR program. 
This publication has received funding from the European Union’s HORIZON Research and Innovation Actions through the project AI4SWEng (Agreement No 101189908).

\bibliographystyle{IEEEtran}
\bibliography{ref}
\end{document}